\documentclass[aps,prx,superscriptaddress,onecolumn]{revtex4}
\usepackage{graphicx} 
\usepackage{amsmath,mathtools,amsfonts,amssymb}
\usepackage{braket}
\usepackage{tikz} 
\definecolor{skyblue}{rgb}{0.447,0.624,0.812} 
\definecolor{scarletred}{rgb}{0.937,0.161,0.161} 
\definecolor{darkgreen}{rgb}{0.0, 0.5, 0.0} 
\definecolor{asparagus}{rgb}{0.53, 0.66, 0.42} 
\usepackage{pgfplots}
\usepackage{ulem}

\pgfplotsset{compat=newest}
\usepgfplotslibrary{external}
\begin{document}

\title{Laser powered dissipative quantum batteries in atom-Cavity QED}

\author{Zamir Bele\~no}
\affiliation{Departamento de F\'{i}sica, Facultad de Ciencias F\'{i}sicas y Matem\'{a}ticas, Universidad de Chile, 837.0415 Santiago, Chile}

\author{Marcelo F. Santos}
\affiliation{Instituto de Física, Universidade Federal do Rio de Janeiro, CP68528, Rio de Janeiro 21941-972, Brasil}

\author{Felipe Barra}
\affiliation{Departamento de F\'{i}sica, Facultad de Ciencias F\'{i}sicas y Matem\'{a}ticas, Universidad de Chile, 837.0415 Santiago, Chile}

\begin{abstract}

The interaction of a three-level atom with the electromagnetic field of a quantum cavity in the presence of a laser field presents a rich behavior that we exploit to discuss two quantum batteries. 
In the first setup, we consider a single three-level atom interacting sequentially with many cavities, each in a thermal state. We show that under this process, the atom converges towards an equilibrium state that displays population inversion. In the second setup, a stream of atoms in a thermal state interacts sequentially with a single cavity initially in a thermal state at the same temperature as the atoms. We show that the cavity's energy increases continuously as the stream of atoms continues to cross, and the cavity does not reach an equilibrium state. However, if we consider the state after many atoms have traveled, the cavity is in an active state that stores energy. The charging process we propose is robust. We discuss its thermodynamics and evaluate the energy supplied by the laser, the energy stored in the battery, and, thus, the device's efficiency.  

\end{abstract}

\maketitle

\section{Introduction}

The quantum nature of light-atom interaction has been studied experimentally in fascinating detail in cavity quantum electrodynamics (CQED) setups, validating the predictions of the Rabi Hamiltonian or simplified models such as the Jaynes-Cummings and Tavis-Cummings models~\cite{HarocheBook}. Quantum thermodynamics uses these developments to propose systems with quantum advantages for energetic or informational purposes~\cite{QTh1,QTh2,QTh3}. For instance, quantum batteries~\cite{Qbat0} store energy and deliver higher power than their classical counterpart~\cite{Qbat1,Qbat2,Qbat3,Qbat4,Qbat5,Qbat6}. A popular system displaying these advantages is the Dicke or Tavis-Cummings battery~\cite{TCBattery}, in which an initial state with some average number of photons charges a given number of qubits, initially in the ground state. Nevertheless, the interaction between the light and the atoms must be turned off at a precise moment to achieve the charged state, hindering its practical implementations. Adiabatic charging protocols~\cite{adiab1} avoid this fine-tuning sacrificing power.

A recent proposal considers a cavity as the battery, which gets charged as it interacts with a stream of atoms prepared in a state with high purity~\cite{micromaserbattery1,micromaserbattery2} that traverses it. The charged battery state reached in this setup is remarkable. However, it is unclear how to evaluate its thermodynamic cost, particularly that of the preparation stage and the stability of the charged state under fluctuations.  

Here, we consider a similar but more robust procedure that can be implemented almost autonomously. The charging protocol does not depend on fine control of the time and considers that all states we need to prepare are thermal. This is important because from a thermodynamic perspective~\cite{resourcetheories}, thermal states are prepared for ``free," and thus, we can evaluate the thermodynamic cost for the charging protocol and evaluate a meaningful efficiency for the device. 
 
We study two setups based on the interaction of a three-level atom interacting with a QED cavity and dispersively with a laser (or maser) field. Remarkably, when the population of the higher energy level is low, the dynamics can be approximated by a qubit in a selective anti-Jaynes-Cummings interaction with a cavity~\cite{mfsantos}. We consider that the initial state of the qubits and cavities is thermal. In the first setup, the atom traverses a sequence of cavities. The effective qubit gets charged in a state with population inversion. The second setup is like the micromaser. A stream of uncorrelated three-level atoms crosses sequentially the cavity. After several atoms have crossed, the cavity is in a charged state. We will find important differences between the setups. The first reaches an equilibrium state, while the second does not (although a quasi-steady state exists in a restricted region of the parameters). Nevertheless, both protocols produce charged states autonomously and robustly when the number of iterations is finite. 

The paper is organized as follows. In section~\ref{sec2}, we present the charging process and a thermodynamic framework that allows us to evaluate our batteries' thermodynamic cost and efficiency. In section~\ref{sec3}, we introduce the Hamiltonian describing the three-level atom interacting with the cavity and the laser. We show that under some conditions stated below, the dynamic reduces to that of a qubit interacting with a cavity with an anti-Jaynes-Cummings coupling~\cite{mfsantos}. In section~\ref{sec4}, we apply the framework of section~\ref{sec2} to the model of section \ref{sec3} when the qubit plays the role of the battery. In section~\ref{sec5}, we do it for the case where the cavity is used as a battery. We end with some conclusions in~\ref{secconclu}.

\section{thermodynamic framework}\label{sec2}
  
 Modeling scattering processes, such as an atom flying through a QED cavity with a time-dependent interaction that neglects the motional degree of freedom, is a common procedure justified by the energy scales involved~\cite{scattwork1,scattwork2}. The conditions under which this modeling is justified for a thermodynamic description of the process are fulfilled in our case.  

The scheme we want to explore is the following: a system $S$, with Hamiltonian $H_S$, interacts sequentially with copies of a system $B$, with Hamiltonian $H_B$, all in the thermal state $\omega_\beta(H_B)\equiv e^{-\beta H_B}/{\rm Tr}[e^{-\beta H_B}]$, at temperature $T=(k_B\beta)^{-1}$. 
Initially, all systems are uncorrelated. 
If $U_{\tau_i}$ is the unitary evolution operator for system $S$ interacting with the $i-$th copy of $B$ for a lapse of time $\tau_i$, the state of $S$ after the $i-$th interaction is
\[
{\mathcal E}_{\tau_i}(\rho_S^{(i-1)})={\rm Tr}_B[U_{\tau_i}\rho_S^{(i-1)}\otimes\omega_\beta(H_B)U_{\tau_i}^\dag]={\rm Tr}_B[U_{\tau_i}\rho_{\rm tot}^{(i-1)}U_{\tau_i}^\dag]={\rm Tr}_B[\rho_{\rm tot}'^{(i)}],
\]
with $\rho_S^{(i-1)}$ the state of $S$ right before the interaction with the $i$-th copy of $B$ begins. 
If the system evolves isolated for a time $t_i$ after the interaction with the $i-$th copy of $B$, the change of state of $S$
is given by the map ${\mathcal U}_{t_i}\circ{\mathcal E}_{\tau_i}$ 
where ${\mathcal U}_{t}(\rho)=e^{-i\frac{t}{\hbar}H_S}\rho e^{i\frac{t}{\hbar}H_S}.$ Thus 
\[
\rho_S^{(i)}\equiv {\mathcal U}_{t_i}\circ{\mathcal E}_{\tau_i}(\rho_S^{(i-1)})
\]
is the state of $S$ prior to the $(i+1)$-th interaction.

In the process $\rho_S^{(i-1)}\to \rho_S^{(i)}$ the change of energy $E={\rm Tr}[H_S\rho_S]$ and von Neumann entropy $S_{\rm vN}=-{\rm Tr}[\rho_S\ln \rho_S]$ can be split as
\begin{align}
\Delta E^{(i)}=W^{(i)}+Q^{(i)}, \label{1law}\\ 
\Delta S_{\rm vN}^{(i)}=\Sigma^{(i)}+\beta Q^{(i)},
\end{align}
with $\Sigma^{(i)}=D(\rho_{\rm tot}'^{(i)}||\mathcal{E}_{\tau_i}(\rho_S^{(i-1)})\otimes\omega_\beta(H_B))\geq 0$ identified with the entropy production, and $D(a||b)\equiv{\rm Tr}[a\ln a-a\ln b]$ the relative entropy between the density matrices $a$ and $b$, always non negative. 
Eq.~\eqref{1law} and the inequality $\Sigma^{(i)}\geq 0$ are like the first and second laws of thermodynamics. 
The heat flowing from $B$ to $S$, $Q^{(i)}={\rm Tr}_B[H_B(\omega_\beta(H_B)-\rho_B'^{(i)})]$, is given by minus the energy change of the $i-$th copy of $B$ in the process, because $\rho_B'^{(i)}={\rm Tr}_S[\rho_{\rm tot}'^{(i)}]$ is the state of the $i-$th copy of $B$ after the interaction with $S$. 
The work done on system $S$ is the energy change of the isolated system composed by $S$ and the $i-$th copy of $B$. 
It can be expressed as
\begin{align}\label{worki}
W^{(i)}=\int_0^{\tau_i} {\rm Tr}[\frac{\partial H_{\rm tot}(t)}{\partial t}\rho_{\rm tot}(t)] dt
\end{align}
where the total Hamiltonian 
\begin{equation}
H_{\rm tot}(t)=H_S+H_B+H_{\rm int}(t)
\end{equation}
is the sum of the free Hamiltonians plus a time-dependent term $H_{\rm int}(t)$ that comprises the interaction, which is turned on at the beginning of the process and off at the end, plus driving terms on system $S$ that are only present during the interaction time.  $\rho_{\rm tot}(t)$ is the solution of $i\hbar\partial_t \rho_{\rm tot}(t)=[H_{\rm tot}(t),\rho_{\rm tot}(t)]$, with initial condition $\rho_{\rm tot}(0)=\rho_{\rm tot}^{(i-1)}$ and thus $\rho_{\rm tot}(\tau_i)=\rho_{\rm tot}'^{(i)}=U_{\tau_i}\rho_{\rm tot}^{(i-1)}U_{\tau_i}^\dag$. Therefore, Eq.~\eqref{worki} represents the work performed by the driving plus switching work~\cite{Philipp}, and can be written as,
$W^{(i)}={\rm Tr}[(H_S+H_B)(\rho_{\rm tot}'^{(i)}-\rho_{\rm tot}^{(i-1)})]$.

This repeated interaction scheme has been intensively studied in the context of open quantum systems~\cite{int-rep}. 
In~\cite{lledo} it was noted that if an hermitian operator $H_S^{*}$ acting on the Hilbert space of $S$ satisfying 
$[U_\tau,H_S^{*}+H_B]=0$ for all $\tau$ exists, then, the state $e^{-\beta H_S^{*}}/Z_S^{*}$ ($Z_S^{*}={\rm Tr}[e^{-\beta H_S^{*}}]$) is an equilibrium state for the map ${\mathcal E}_\tau$. The main properties are
\begin{itemize}
\item For any $\tau_i$, ${\mathcal E}_{\tau_i}(e^{-\beta H_S^{*}}/Z_S^{*})=e^{-\beta H_S^{*}}/Z_S^{*}$,  which is diagonal in the energy eigenbasis of $S$ if $[H_S^*,H_S]=0$. 
\item For any pair of sequences $\tau_1,\tau_2,\ldots$, and $t_1,t_2,\ldots$, the iterative process converges to the equilibrium state, i.e., 
\begin{equation}
 \rho_S^{(0)}\xrightarrow[]{{{\mathcal U}_{t_1}\circ\mathcal E}_{\tau_1}} \rho_S^{(1)}\xrightarrow[]{{\mathcal U}_{t_2}\circ{\mathcal E}_{\tau_2}}\cdots\to\rho_S^{(i-1)}\xrightarrow[]{{\mathcal U}_{t_i}\circ{\mathcal E}_{\tau_i}} \rho_S^{(i)}\xrightarrow[i\to \infty]{}\frac{e^{-\beta H_S^{*}}}{Z_S^{*}},\quad \forall  \rho_S^{(0)}.
\label{itpro1}
\end{equation}  
\item Once the invariant state is reached, keeping the charged state has no thermodynamic cost. Thus, the total work performed in the process that charged the system from the initial state $\rho_S^{(0)}$ is~\cite{barra2019}
\begin{equation}\label{totwork}
W_{\rm tot}=\sum_i W^{(i)}={\rm Tr}\left[(H_S-H_S^*)\left(\frac{e^{-\beta H_S^{*}}}{Z_S^{*}}-\rho_S^{(0)}\right)\right]
\end{equation}
 \end{itemize}
 
 Furthermore, in~\cite{barra2019,barra2022}, it was shown that equilibrium states with population inversion can be achieved. The key to obtaining such equilibrium states is to design a setup such that if the system Hamiltonian is $H_S=\sum_i E_i\ket{E_i}\bra{E_i}$ with $E_1\leq E_2\leq \cdots$, the evolution operator $U_\tau$ satisfies $[U_\tau,H_S^{*}+H_B]=0$ with a hermitian $H_S^{*}=\sum_i E_i^*\ket{E_i}\bra{E_i}$ where at least for some $i:$ $E_i^*>E_{i+1}^*$. In this way $e^{-\beta H_S^{*}}/Z_S^{*}$ has population inversion. One way of characterizing the charge of the system is with its ergotropy~\cite{ergotropy}, which is the maximum energy that can be extracted from this state in a unitary process.  
 The ergotropy of the state $e^{-\beta H_S^{*}}/Z_S^{*}$ for a system with Hamiltonian $H_S$ commuting with $H_S^*$ is given by~\cite{barra2019}
\begin{equation}\label{ergoeq}
{\mathcal W}=\sum_i(E_{p_i}-E_i)\frac{e^{-\beta E^*_{p_i}}}{Z_S^*}.
\end{equation}
The state of the system after ergotropy extraction is $\sigma_{\omega_\beta(H_S^*)}=\sum_i\frac{e^{-\beta E^*_{p_i}}}{Z^*_S}\ket{E_i}\bra{E_i}$, which is a passive state~\cite{passive} meaning that its ergotropy vanishes. 
In these two expressions $p$ is a permutation of the indices $\{1,2\ldots\}$ such that $E_1^*\leq E_2^*\leq \cdots$.
According to the second law, the ergotropy is always smaller than the work cost $W_{\rm tot}$ for the process that starts with $\rho_S^{(0)}=\sigma_{\omega_\beta(H_S^*)}$.
 
The advantage of this charging protocol is that it is robust in the sense that we do not need a precise time control for $\{\tau_i\}$ and $\{t_i\}$, and that, due to the contracting character of the map $\mathcal{E}_\tau$, any ``deviation" in the charging process is corrected.   

From a thermodynamics perspective, given a bath with inverse temperature $\beta$, a system with Hamiltonian $H_S$ can be prepared in the state $\omega_\beta(H_S)$ for ``free"~\cite{resourcetheories}. Therefore, apart from considering the copies of $B$ in their thermal state, we also consider the initial state of $S$, for the process \eqref{itpro1}, in the thermal state $\rho_S^{(0)}=\omega_\beta(H_S)$ although other initial conditions may be considered. 

We will consider this scheme in the context of a three-level atom driven by a laser field while interacting with a quantum electromagnetic cavity. These determine $H_{\rm int}(t)$. 
We apply this in two scenarios. First, the atom is the system $S$ that starts in a thermal state and crosses a sequence of thermal electromagnetic cavities, playing the role of $B$. 
We analyze the asymptotic state of the atom and show that it reaches a state with population inversion. Then, we reverse the setup and 
consider a stream of thermal atoms (playing the role of $B$) that crosses a single cavity corresponding to system $S$, which is initially in a thermal state. We analyze the time-dependent state of the cavity and discuss its energetics.

\section{model}
\label{sec3}

This section describes the dynamics of the driven atom interacting with the cavity. As shown in~\cite{mfsantos}, there is an interesting regime in which the dynamic is well approximated by a qubit (two-level system) that couples to the cavity with an anti-Jaynes-Cummings term.    

Before entering the cavity and after leaving the cavity, the three-level atom is free, and its Hamiltonian is $H_{\rm atom}=\hbar \omega_g\ket{g}\bra{g}+\hbar \omega_e\ket{e}\bra{e}+\hbar \omega_h\ket{h}\bra{h}$, with $\omega_g<\omega_e<\omega_h$.
The total Hamiltonian is
\begin{equation}
H=H_{\rm atom}+H_{\rm cavity}+H_{\rm int}(t)
\end{equation}
with 
\begin{align}
H_{\rm cavity}&= \hbar \omega a^\dag a,\\
H_{\rm int}(t)&=\left\{\hbar\left(\Omega_L\ket{h}\bra{g}e^{-i\omega_L t}+g\ket{h}\bra{e}a+\textrm{H.C}\right)+\hbar\epsilon\ket{g}\bra{g}\right\}\chi(t),
\end{align}
and $\chi(t)=1$ if at time $t$ the atom is inside the cavity and $\chi(t)=0$, elsewhere. The first term in $H_{\rm int}(t)$ corresponds to the laser of frequency $\omega_L$ driving dispersively the transition from level $\ket{g}$ to level $\ket{h}$ of the atom, with coupling frequency $\Omega_L$. The second term corresponds to the interaction between the atom and the cavity mode of frequency $\omega$, which couples level $\ket{e}$ to level $\ket{h}$, with coupling frequency $g$. $a$ and $a^\dag$ are the destruction and creator operators of the mode. The last term modifies the atom energy levels as a DC Stark effect can do. The detuning is $\delta=\omega_h-\omega_e-\omega$.

As a quick remark, we note from $H_{\rm int}(t)$ and Eq.~\eqref{worki} that the work done by the driving laser (or maser) has an order of magnitude determined by ($\sim \hbar\Omega_L\omega_L\tau$) and the work done switching on and off the interaction by ($\sim \hbar\Omega_L$ or $\sim \hbar g$). Since $ \tau \sim 1/\Omega_L,1/g$ and $\omega_L \gg \Omega_L,g$, the laser does most of the work 
\footnote{Considering the order of magnitudes given at the end of the conclusion, for a setup in which a stream of thermal atoms interact with a maser and a microwave cavity, the work performed by the maser is $\sim 10^6$ larger than the switching work. } .

Let us now analyze the model in the interaction picture,
$\ket{\tilde{\Phi}(t)}\equiv e^{iH_0t/\hbar}\ket{\Phi(t)}$, where $H_0=H_{\rm atom}+H_{\rm cavity}$, which satisfies
\begin{equation}
\label{schoeq}
i\hbar\partial_t\ket{\tilde{\Phi}(t)}=\tilde{H}_{\rm int}(t)\ket{\tilde{\Phi}(t)}
\end{equation}
with
\begin{equation}
\tilde{H}_{\rm int}(t)=\hbar\left(\Omega_L\ket{h}\bra{g}e^{i(\omega_h-\omega_g)t}e^{-i\omega_L t}+g\ket{h}\bra{e}e^{i(\omega_h-\omega_e)t}e^{-i\omega t}a+\textrm{H.C}\right)+\hbar\epsilon\ket{g}\bra{g}.
\end{equation}

Taking $\delta=\omega_h-\omega_g-\omega_L=\omega_h-\omega_e-\omega$ we have
\begin{equation}
\label{prl1}
\tilde{H}_{\rm int}(t)=\hbar\left(\Omega_L\ket{h}\bra{g}e^{i\delta t}+g\ket{h}\bra{e}e^{i\delta t}a+\textrm{H.C}\right)+\hbar\epsilon\ket{g}\bra{g}
\end{equation}
and, considering $\ket{\tilde{\Phi}(t)}=c_g(t)\ket{g,n}+c_e(t)\ket{e,n+1}+c_h(t)\ket{h,n}$, with $\{\ket{n}\}$ the eigenstates of $a^\dag a$, we get the closed system of equations 
\begin{align}
i\dot c_g&=\epsilon c_g+\Omega_L^*e^{-i\delta t}c_h\label{cgdot}\\
i\dot c_e&=g^*\sqrt{n+1}e^{-i\delta t}c_h\label{cedot}\\
i\dot c_h&=\Omega_L e^{i\delta t} c_g+g\sqrt{n+1}e^{i\delta t}c_e \label{chdot}.
\end{align}

If $\delta\gg |\Omega_L|$, $\delta\gg |g|$ and initially $c_h\approx 0$ one can adiabatically eliminate level $\ket{h}$ and obtain that $\ket{\tilde{\Phi}(t)}=c_g(t)\ket{g,n}+c_e(t)\ket{e,n+1}$ evolves 
with a time-independent Hamiltonian in the interaction picture given by
\begin{align}
\tilde{V}_\textrm{eff}(t)=\tilde{V}_\textrm{eff}=-\hbar\left([-\epsilon+\frac{|\Omega_L|^2}{\delta}]\ket{g}\bra{g}+\frac{|g|^2}{\delta}\ket{e}\bra{e}a^\dag a+\frac{\Omega_Lg^*}{\delta}\ket{g}\bra{e}a+\textrm{H.C}\right).
\label{prl2}
\end{align}

Thus, in this regime, we can consider that the atom is a two-level system with Hamiltonian $H_{\rm qubit}=\hbar \omega_g\ket{g}\bra{g}+\hbar \omega_e\ket{e}\bra{e}$ that evolves with the cavity according to $i\hbar\partial_t\ket{\tilde{\Phi}(t)}=\tilde{V}_\textrm{eff}(t)\ket{\tilde{\Phi}(t)}$ in the interaction picture. The state in the Schr\"odinger picture is $e^{-iH_0t/\hbar}\ket{\tilde{\Phi}(t)}$ with 
\begin{equation}
\label{H0}
H_0=\hbar \omega_g\ket{g}\bra{g}+\hbar \omega_e\ket{e}\bra{e}+\hbar \omega a^\dag a.
\end{equation}
Note that the coupling in Eq.~\eqref{prl2} has the anti-Jaynes-Cummings structure.

From Eq.~\eqref{prl2}, we can compute the evolution operator in the interaction picture. For this, it is convenient to define $N_0$ by the relation 
\begin{equation}\label{epsilon}
\epsilon=\frac{|\Omega_L|^2}{\delta}-\frac{|g|^2}{\delta}(N_0+1),
\end{equation}
and the auxiliary quantities 
\begin{equation}\label{Deltan}
\Delta_n\equiv \frac{|g|^2}{\delta}(n-N_0),
\end{equation} 
\begin{equation}\label{Gn}
G_n\equiv \frac{|\Omega_L g^*|}{\delta}\sqrt{n+1},
\end{equation} 
and  
\begin{equation}\label{relutil}
\Omega_n^2\equiv\Delta_n^2/4+G_n^2.
\end{equation} 
In terms of these, the evolution operator in the interaction picture is
\begin{align}
U_I(t)\equiv e^{-it\tilde{V}_{\rm eff}/\hbar}=\ket{e,0}\bra{e,0}+\sum_{n=0}^\infty\left\{\left(\cos(\Omega_n t)+i\frac{\Delta_n}{2\Omega_n}\sin(\Omega_n t)\right)\ket{g,n}\bra{g,n}\nonumber+\right.\\ 
\left.
\left(\cos(\Omega_n t)-i\frac{\Delta_n}{2\Omega_n}\sin(\Omega_n t)\right)\ket{e,n+1}\bra{e,n+1}-
\frac{iG_n}{\Omega_n}\sin(\Omega_n t)(\ket{g,n}\bra{e,n+1}+\ket{e,n+1}\bra{g,n})\right\}.
\label{UItau}
\end{align} 
We have considered the eigenbasis of the number operator $a^\dag a\ket{n}=n\ket{n}$.
With these quantities at hand, we are ready to study our two proposed quantum batteries.



\section{Dissipative charging of the qubit}\label{sec4}

In this subsection, we consider the charging process discussed in section~\ref{sec2} for an atom crossing many thermal cavities in series. The interaction between the atom and the cavity in the regime discussed in section~\ref{sec3} is equivalent to an effective qubit-cavity system evolving with $U_\tau=e^{i\tau H_0}U_I(\tau)$ where $U_I(t)$ is given in Eq.~\eqref{UItau} and $H_0$ in Eq.~\eqref{H0}.
Thus, the qubit is the system $S$, and the cavities play the role of $B$. Any time the atom crosses a thermal cavity in a time interval $\tau$, the qubit changes its state from $\rho_{\rm qubit}$ to 
\[
\mathcal{E}_\tau(\rho_{\rm qubit})={\rm Tr}_{\rm cavity}[U_\tau\rho_{\rm qubit}\otimes\omega_\beta(H_{\rm cavity})U^\dag_\tau].
\]

As follows from section \ref{sec2}, the charging process reaches an equilibrium state $e^{-\beta H_{\rm qubit}^{*}}/Z_{\rm qubit}^{*}$ for the qubit if $[U_\tau,H_{\rm qubit}^{*}+H_{\rm cavity}]=0$ or if $[\tilde{V}_{\rm eff},H_{\rm qubit}^*+H_{\rm cavity}]=0$ where $\tilde{V}_{\rm eff}$ is given in Eq.~\eqref{prl2} and $[H_{\rm qubit}^*,H_{\rm qubit}]=0$.
A direct computation shows that $H_{\rm qubit}^{*}=\hbar\omega\ket{g}\bra{g}$ satisfies these requeriments. 

This means that 
\[
\frac{e^{-\beta H_{\rm qubit}^{*}}}{Z_{\rm qubit}^{*}}=\frac{e^{-\beta\hbar\omega}}{1+e^{-\beta\hbar\omega}}\ket{g}\bra{g}+\frac{1}{1+e^{-\beta\hbar\omega}}\ket{e}\bra{e}=p_{g}^{(\infty)}\ket{g}\bra{g}+p_{e}^{(\infty)}\ket{e}\bra{e}
\] 
is an equilibrium state for the process \eqref{itpro1}. Since the population in the excited state is higher than in the ground state, we say that the battery is charged. We quantify its charge with the ergotropy, which we evaluate using Eq.~\eqref{ergoeq} and obtain
\[
\mathcal{W}=\hbar(\omega_e-\omega_g)(p_{e}^{(\infty)}-p_{g}^{(\infty)})
=\hbar(\omega_e-\omega_g)\tanh\frac{\beta\hbar\omega}{2}
\]
Once the ergotropy is extracted, the battery (qubit) is left in the passive state $\sigma_{\omega_\beta}(H_{\rm qubit}^*)=p_e^{(\infty)}\ket{g}\bra{g}+p_g^{(\infty)}\ket{e}\bra{e}$

According to Eq.~\eqref{totwork}, the total work performed in the charging process $\rho_S^{(0)}\to \omega_\beta(H_{\rm qubit}^{*})$ is
\[
W^{(\rho_S^{(0)})}_{\rm tot}={\rm Tr}[(\hbar\omega_e\ket{e}\bra{e}+\hbar(\omega_g-\omega)\ket{g}\bra{g})(\omega_\beta(H_{\rm qubit}^{*})-\rho_S^{(0)})].
\]
We indicate with a superscript the initial state to distinguish two cases in what follows.
If $\rho_S^{(0)}=\omega_\beta(H_{\rm qubit})$
then 
\[
W^{(\omega_\beta)}_{\rm tot}=\hbar\left(\frac{(\omega_g-\omega)e^{-\beta\hbar\omega}+\omega_e}{1+e^{-\beta\hbar\omega}}-\frac{(\omega_g-\omega)+\omega_ee^{-\beta\hbar(\omega_e-\omega_g)}}{e^{-\beta\hbar(\omega_e-\omega_g)}+1}\right).
\]
If $\rho_S^{(0)}=\sigma_{\omega_\beta}(H_{\rm qubit}^*)$ is the passive state after ergotropy extraction then 
\[
W^{(\sigma_{\omega_\beta})}_{\rm tot}=\hbar(\omega_e-\omega_g+\omega)\tanh\frac{\beta\hbar\omega}{2}=
\hbar\omega_L\tanh\frac{\beta\hbar\omega}{2},
\]
manifesting the role of the laser powering the battery. In the last equation, we have used the equalities $\delta=\omega_h-\omega_g-\omega_L=\omega_h-\omega_e-\omega$, that implies $\omega_L=\omega_e-\omega_g+\omega,$ and thus $\omega_L>\omega$.

We define the efficiency of the device in the corresponding charging process by
\[
\eta^{(\rho_S^{(0)})}=\frac{\mathcal{W}}{W^{(\rho_S^{(0)})}_{\rm tot}}.
\]
Note that $\eta^{(\sigma_{\omega_\beta})}\leq 1$ due to the second law. 

So, if $\omega>\omega_e-\omega_g \Leftrightarrow 2\omega>\omega_L$, the efficiency $\eta^{(\omega_\beta)}$ monotonically increase from its low temperature limit ($\beta\to \infty$)
\begin{equation}
\label{etalow}
\lim_{\beta\to \infty}\eta^{(\omega_\beta)}=\frac{(\omega_e-\omega_g)}{(\omega_e-\omega_g+\omega)}=
\frac{(\omega_L-\omega)}{\omega_L}
\end{equation}
to its high-temperature limit ($\beta\to 0$)
\[
\lim_{\beta\to 0}\eta^{(\omega_\beta)}=\frac{2(\omega_e-\omega_g)\omega}{ [\omega_e-\omega_g+\omega]^2}=
\frac{2(\omega_L-\omega)\omega}{\omega_L^2},
\]

On the other hand, considering the process that recharges the qubit after ergotropy extraction, the efficiency
\[
\eta^{(\sigma_{\omega_\beta})}=\frac{\omega_e-\omega_g}{\omega_e-\omega_g+\omega}=
\frac{\omega_L-\omega}{\omega_L},\,\forall \beta
\]
equals the efficiency Eq.~\eqref{etalow} for an initial thermal qubit in the low-temperature limit. Therefore, if $2\omega>\omega_L$, it is better to use a thermal qubit rather than recycling a passive qubit. If 
$2\omega<\omega_L$ then $\eta^{(\sigma_{\omega_\beta})}\geq \eta^{(\omega_\beta)},\forall \beta>0$, and therefore in this parameter regime it is better to recycle the used qubit. 

 \begin{figure}[t] 
\centering
\includegraphics[width=0.4\textwidth]{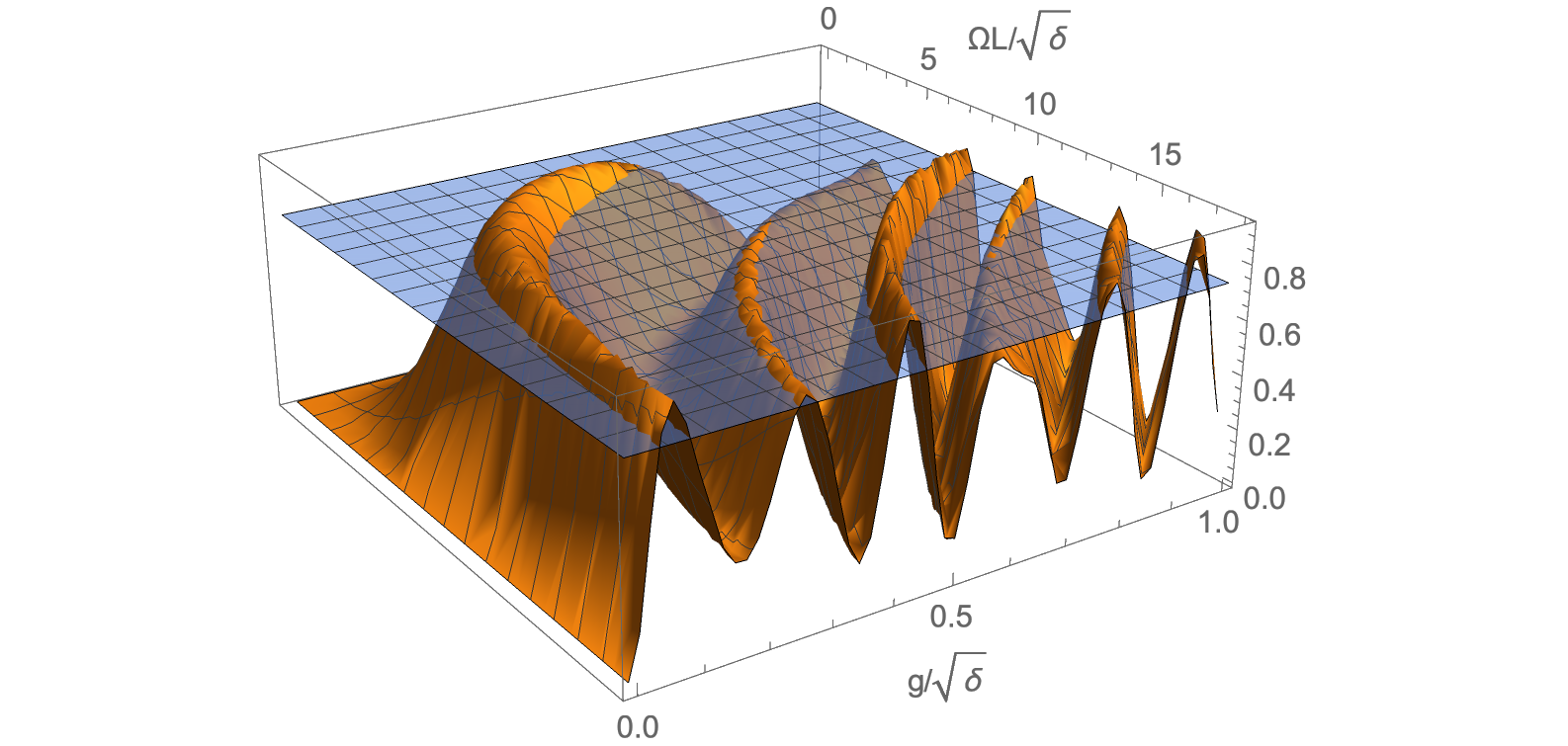}
    \caption{Plot of $A_\tau$ and $1/(1+e^{-\beta \omega})$ as a function of $\Omega_L/\sqrt{\delta},g/\sqrt{\delta}$. Other parameters are $\beta\omega=\tau=1$ and $N_0=6.5$.}
    \label{figArate}
\end{figure}

To analyze the time evolution of the battery during the charging process, we evaluate the map,
\[
\mathcal{E}_\tau(\rho_{\rm qubit})=\sum_{n=0}^\infty \frac{e^{-\beta\hbar\omega n}}{Z_{\rm cavity}}{\rm Tr}_{\rm cavity}[U_\tau\rho_{\rm qubit}\otimes \ket{n}\bra{n}U^\dag_\tau]
\]
with the partition function of the cavity given by
\[
Z_{\rm cavity}=\sum_{n=0}^\infty  e^{-\beta\hbar\omega n}=
\frac{e^{\beta\hbar\omega/2}}{2\sinh(\beta\hbar\omega/2)}.
\]

Considering a diagonal state $\rho_{\rm qubit}=p_g\ket{g}\bra{g}+p_e\ket{e}\bra{e}$, one has
\[
{\rm Tr}_{\rm cavity}[U_\tau\rho_{\rm qubit}\otimes \ket{n}\bra{n}U^\dag_\tau]
={\rm Tr}_{\rm cavity}[e^{i\tau H_0}U_I(\tau)\rho_{\rm qubit}\otimes \ket{n}\bra{n}U^\dag_I(\tau)e^{-i\tau H_0}],
\]
where we have expressed the unitary operator in the interaction picture $U_\tau=e^{i\tau H_0/\hbar}U_I(\tau)$, with $U_I(\tau)$ given in Eq.~\eqref{UItau} and $H_0$ in Eq.~\eqref{H0}. After some algebra, one obtains
\begin{align}
\rho_{\rm qubit}(\tau)=\left(p_g(1-A_\tau)+p_ee^{-\beta\hbar\omega}A_\tau\right)\ket{g}\bra{g}+
\left(p_e(1-A_\tau e^{-\beta\hbar\omega})+p_gA_\tau\right)\ket{e}\bra{e}
\end{align}
with 
\begin{align}
A_\tau\equiv \sum_{n=0}^\infty  \frac{G_n^2}{\Omega_n^2}\frac{e^{-\beta\hbar\omega n}}{Z_{\rm cavity}}\sin^2(\Omega_{n} \tau).
\end{align}
Note that $0<A_\tau<1$. So if $\rho_{\rm qubit}$ is diagonal in the energy basis, $\rho_{\rm qubit}(\tau)$ is also diagonal. Note that diagonal states are invariant under the free evolution ${\mathcal U}_t$, and thus, if the qubit starts in a thermal state, it stays diagonal at every stage of the iteration process \eqref{itpro1} and the free evolutions between iterations are irrelevant.
Denoting $\rho^{(i)}_{\rm qubit}=p_g^{(i)}\ket{g}\bra{g}+p_e^{(i)}\ket{e}\bra{e}$, the state of the qubit after the $i-$th interaction with the cavity, one can write the map $\rho^{(i-1)}_{\rm qubit}\to \rho^{(i)}_{\rm qubit}$ in terms of a transition probability matrix,
\begin{align}
\left(\begin{array}{c}
p_g^{(i)}\\
p_e^{(i)}
\end{array}\right)=
\left(\begin{array}{cc}
(1-A_{\tau_i})&A_{\tau_i}e^{-\beta\hbar\omega}\\
A_{\tau_i}&1-A_{\tau_i}e^{-\beta\hbar\omega}
\end{array}\right)
\left(\begin{array}{c}
p_g^{(i-1)}\\
p_e^{(i-1)}
\end{array}\right)
\label{qubitmatrix}
\end{align}
which for any sequence $\tau_1,\tau_2,\ldots$ it converges to the stationary state solution~\footnote{It is simple to show that $p^{(i+1)}_{g}-p^{(\infty)}_{g}=(1-A_{\tau_i})[p^{(i)}_{g}-p^{(\infty)}_{g}]<p^{(i)}_{g}-p^{(\infty)}_{g}$}
\[
p_e^{(\infty)}=\frac{1}{1+e^{-\beta\hbar\omega}},\quad p_g^{(\infty)}=\frac{e^{-\beta\hbar\omega}}{1+e^{-\beta\hbar\omega}}.
\]
This asymptotic state for the atom shows population inversion and is, in fact $e^{-\beta H_{\rm qubit}^{*}}/Z_{\rm qubit}^{*}$ with $H_{\rm qubit}^{*}=\hbar\omega\ket{g}\bra{g}$ as anticipated. 

We can also evaluate the time scale at which the charged state is achieved. 
The eigenvalues of the matrix in Eq.~\eqref{qubitmatrix} are $1$, (i.e. there is an invariant state) and $1-A_{\tau_i}(e^{-\beta \hbar\omega}+1)$, so population inversion is achieved at a rate $-\ln |1-A_{\tau_i}(e^{-\beta \hbar\omega}+1)|$ and so to achieve a fast charging rate we need $A_{\tau}\to1/(e^{-\beta \hbar\omega}+1)$.
In figure~\ref{figArate}, we plot $A_\tau$ as a function of $\Omega_L/\sqrt{\delta},g/\sqrt{\delta}$, showing that the optimal rate can be achieved.
Note that to increase the ratio $G_n/\Omega_n$ and thus $A_\tau$ one must consider $|\Omega_L|\gg |g|.$

\section{Dissipative charging of a cavity: a micromaser quantum battery}\label{sec5}

We now consider the cavity as the system $S$ to be charged and the qubits as the thermal systems $B$ assisting the charging process. 
If we look for an operator $H_{\rm cavity}^{*}$ for the cavity, such that $[U_\tau,H_{\rm cavity}^{*}+H_{\rm qubit}]=0$ with $U_\tau=e^{-iH_0\tau/\hbar}U_I(\tau)$ in Eqs.~\eqref{H0} and \eqref{UItau} and $H_{\rm qubit}=\hbar\omega_g\ket{g}\bra{g}+\hbar\omega_e\ket{e}\bra{e}$, with the intention of finding an equilibrium state$\sim e^{-\beta H_{\rm cavity}^{*}}$, we find $H_{\rm cavity}^{*}=-\hbar(\omega_e-\omega_g)a^\dag a$.
Since $\omega_e>\omega_g,$ $e^{-\beta H_{\rm cavity}^{*}}$ does not have a trace, and we conclude that there is no equilibrium state for the cavity. To confirm this, we study the time-dependent process.

Consider the cavity in the initial state $\rho_{\rm cavity}=\sum_{n=0}p_n\ket{n}\bra{n}.$ The state of the cavity after the crossing of a thermal atom is
\[
\rho_{\rm cavity}(\tau)={\rm Tr}_{\rm qubit}[U_\tau\omega_\beta(H_{\rm qubit})\otimes\rho_{\rm cavity}U_\tau^\dag]=\sum_{n=0}^\infty p_n{\rm Tr}_{\rm qubit}[U_\tau\omega_\beta(H_{\rm qubit})\otimes \ket{n}\bra{n}U_\tau^\dag]
\]
with $\omega_\beta(H_{\rm qubit})=p^{(\rm th)}_g\ket{g}\bra{g}+p^{(\rm th)}_e\ket{e}\bra{e}$ and $p^{(\rm th)}_e/p^{(\rm th)}_g=e^{-\beta\hbar(\omega_e-\omega_g)}$, the thermal state for the qubit.
After some algebra, it is possible to show that $\rho_{\rm cavity}(\tau)$ is also diagonal in the number basis and thus the map 
\[
\rho_{\rm cavity}^{(i-1)}=\sum_{n=0}p_n^{(i-1)}\ket{n}\bra{n}\to \rho_{\rm cavity}^{(i)}=\sum_{n=0}p_n^{(i)}\ket{n}\bra{n}\] 
is determined by the relation
\begin{equation}
p_n^{(i)}=p_{n-1}^{(i-1)} \left[p^{(\rm th)}_g\frac{G_{n-1}^2}{\Omega_{n-1}^2}\sin^2(\Omega_{n-1} \tau_i)\right]+
p_n^{(i-1)} d_n^{(i)}
+p_{n+1}^{(i-1)} \left[p^{(\rm th)}_e\frac{G_{n}^2}{\Omega_{n}^2}\sin^2(\Omega_{n} \tau_i)\right]
\label{tridiag}
\end{equation}
with
\[
d_n^{(i)}=p^{(\rm th)}_g\left(\cos^2(\Omega_n \tau_i)+\frac{\Delta_n^2}{4\Omega_n^2}\sin^2(\Omega_n \tau_i)\right)+
p^{(\rm th)}_e\left(\cos^2(\Omega_{n-1} \tau_i)+\frac{\Delta_{n-1}^2}{4\Omega_{n-1}^2}\sin^2(\Omega_{n-1} \tau_i)\right).
\]
Eq.~\eqref{tridiag} can be written as ${\bf p}^{(i)}={\bf M}_{\tau_i}{\bf p}^{(i-1)}$ in terms of the probability vectors ${\bf p}^{(i)}=\{p_n^{(i)}\}_{n\geq 0}$ and a tridiagonal matrix ${\bf M}_{\tau_i}$ satisfying  $\sum_m [{\bf M}_\tau]_{mn}=1, \forall n,\tau$ implying $\sum_n p_n^{(i)}=1,\forall i$. 

\begin{figure}[h] 
\centering
\includegraphics[width=0.4\textwidth]{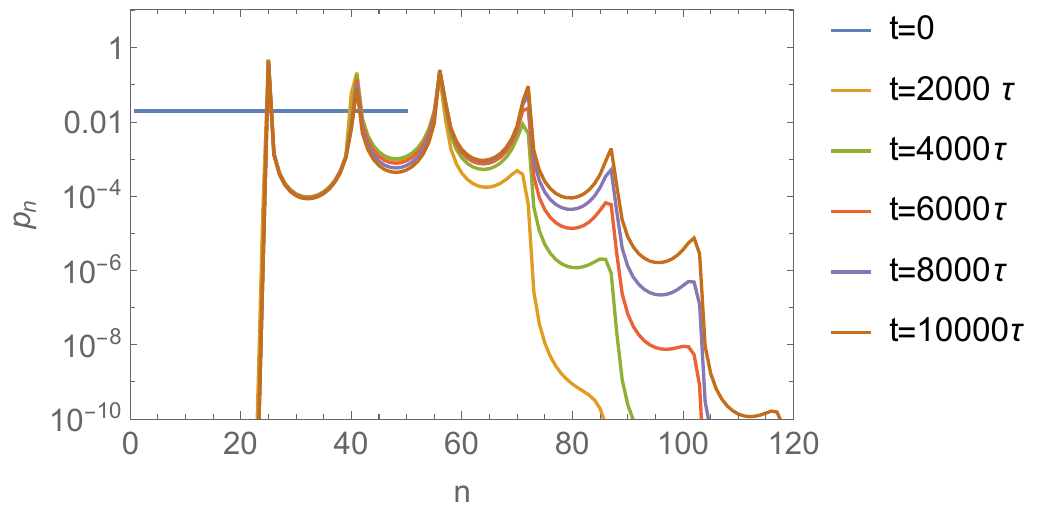}
    \caption{Plot of $p_n^{(i)}$ versus $n$ for $i=j\times 10^{3}$ and $j=2,4,6,8,10$. Parameters $\tau=1.0, N_0=10.5, \Omega_L=1, \delta=8, \omega_e-\omega_g=1.3,\beta=10,g=1.8.$. The initial condition is uniform between 0 and 50 and is also depicted in the plot.}
    \label{figuralog}
\end{figure}

It can be directly checked, using Eq.~\eqref{relutil}, that an invariant vector ${\bf p}^{(\infty)}$ satisfying ${\bf p}^{(\infty)}={\bf M}_\tau{\bf p}^{(\infty)},\forall \tau$ has components of the form $p_n^{(\infty)}\sim (p^{(\rm th)}_g/p^{(\rm th)}_e)^n$ and thus not normalizable ($p^{(\rm th)}_g\geq p^{(\rm th)}_e$ for thermal states). 
In fact, there is no steady-state solution for all values of the parameters. In Figure~\ref{figuralog}, we plot $p_n$ as a function of $n$ at several times $t_k=k\tau$, and we see that the probability shifts towards higher values of $n$. 

\subsection*{limit of high selectivity} 

Tuning $\epsilon$ (see Eq.~\eqref{epsilon}) to obtain an integer value $N_0$ and reducing the laser intensity so that it acts as a weak perturbation $g\gg \Omega_L$, we found an interesting regime for our battery. 

In fact, noticing that $g\gg \Omega_L$ implies $\Omega_n\approx\Delta_n/2\gg G_n$ for $n\neq N_0$, we can neglect  $G_n/\Omega_n, n\neq N_0$ in Eq.~\eqref{UItau}. See Eqs.~\eqref{Deltan},\eqref{Gn} and \eqref{relutil}.
Moreover, $\Delta_n=0$ for $n=N_0$ implies $\Omega_{N_0}=G_{N_0}$ and thus,
the unitary operator $U_I(t)$ becomes
\begin{small}
\begin{align}
U_I(t)&=\cos(G_{N_0} t)\left(\ket{g,N_0}\bra{g,N_0}+
\ket{e,N_0+1}\bra{e,N_0+1}\right)-
i\sin(G_{N_0} t)(\ket{g,N_0}\bra{e,N_0+1}+\ket{e,N_0+1}\bra{g,N_0})+\notag \\
&\sum_n\left\{
e^{i\Delta_n/2 t}\ket{g,n}\bra{g,n}+
e^{-i\Delta_n/2 t}\ket{e,n+1}\bra{e,n+1}
\right\}+\ket{e,0}\bra{e,0},\label{Uselect}
\end{align}
\end{small}
so there is a closed oscillatory dynamics in $\{\ket{e,N_0+1},\ket{g,N_0}\}$ that leaves the other levels invariant (except for a phase). This is the limit of high selectivity~\cite{mfsantos} in which an atom entering the cavity in its ground state $\ket{g}$ can make a transition to $\ket{e}$ only when the cavity has $N_0$ photons. This selective dynamics has been shown to allow for unitary manipulations of the quantum states of both the harmonic oscillator~\cite{marcelo4} or the atom, including a full unitary charging of the latter~\cite{marcelo1}. More relevant in our context is the fact that, in this regime, the equilibrium state $\omega_\beta(H^{*'}_{\rm cavity})$ with $H^{*'}_{\rm cavity}=-\hbar(\omega_e-\omega_g)[N_0\ket{N_0}\bra{N_0}+(N_0+1)\ket{N_0+1}\bra{N_0+1}]$ involving population inversion on the two-dimensional sector of the cavity $\ket{N_0},\ket{N_0+1}$ is achieved. This is obtained by solving $[U_I(t),H_{\rm qubit}+H^{*'}_{\rm cavity}]=0$ with $U_I(t)$ given in Eq.~\eqref{Uselect}, but we can obtain it by analyzing the time-dependent process.

The matrix ${\bf M}_\tau$ in Eq.~\eqref{tridiag} for $g\gg \Omega_L$ changes from tridiagonal to diagonal with $d_n=p^{(\rm th)}_g+p^{(\rm th)}_e=1 \forall n\neq \{N_0,N_0+1\}$, i.e., in this approximation ${\bf M}_\tau$ is the identity outside the sector $\{N_0,N_0+1\}$. The dynamics takes place only in the resonant sector $\{N_0,N_0+1\}$, where 
\begin{align}
d_{N_0}^{(i)}
\approx p^{(\rm th)}_g\cos^2(G_{N_0} \tau_i)+p^{(\rm th)}_e\\
d_{N_0+1}^{(i)}\approx p^{(\rm th)}_g+p^{(\rm th)}_e\cos^2(G_{N_0} \tau_i)
\end{align}
and thus, 
using $p^{(\rm th)}_e=1-p^{(\rm th)}_g$, we obtain 
\begin{align}\label{mcavity}
\left(\begin{array}{c}
p_{N_0}^{(i)}\\
p_{N_0+1}^{(i)}
\end{array}\right)=
\left(\begin{array}{cc}
1-p^{(\rm th)}_g\sin^2(G_{N_0} \tau_i)&(1-p^{(\rm th)}_g)\sin^2(G_{N_0} \tau_i)\\
p^{(\rm th)}_g\sin^2(G_{N_0} \tau_i)&1-(1-p^{(\rm th)}_g)\sin^2(G_{N_0} \tau_i)
\end{array}\right)
\left(\begin{array}{c}
p_{N_0}^{(i-1)}\\
p_{N_0+1}^{(i-1)}
\end{array}\right).
\end{align}
If the initial probability is normalized in this sector, i.e., $p_{N_0}^{(0)}+p_{N_0+1}^{(0)}=1$, 
the occupations of $N_0$ and $N_0+1$ will converge to the invariant distribution
\begin{equation}\label{chselect}
\left(\begin{array}{c}
p_{N_0}^{(\infty)}\\
p_{N_0+1}^{(\infty)}
\end{array}\right)=\left(\begin{array}{c}
1-p^{(\rm th)}_g\\
p^{(\rm th)}_g
\end{array}\right),
\end{equation}
which corresponds to $\omega_\beta(H^{*'}_{\rm cavity})$. If the initial probability $p_n^{(0)}$ is distributed over many values of $n$, for instance, a thermal distribution for the photon number, then under iteration $p_n^{(i)}$ for $n\notin \{N_0,N_0+1\}$ stay constant and $p_{N_0+1}^{(\infty)}=e^{\beta \hbar(\omega_e-\omega_g)}p_{N_0}^{(\infty)}$
with $p_{N_0+1}^{(\infty)}+p_{N_0}^{(\infty)}=p_{N_0+1}^{(0)}+p_{N_0}^{(0)}$. In any case, 
in the limit of high selectivity, the population of the resonant sector is inverted, and the rest remains invariant as time progresses. The work Eq.~\eqref{totwork} in this charging process is 
\[
W_{\rm tot}=\hbar(\omega-\omega_g+\omega_e)[(N_0+1)(p_{N_0+1}^{(\infty)}-p_{N_0+1}^{(0)})+N_0(p_{N_0}^{(\infty)}-p_{N_0}^{(0)})]=\hbar\omega_L(p_{N_0+1}^{(\infty)}-p_{N_0+1}^{(0)}).
\]

The ergotropy, Eq.~\eqref{ergoeq}, of the charged state is ${\mathcal W}=\hbar\omega(p_{N_0+1}^{(\infty)}-p_{N_0}^{(\infty)})$, and the efficiency 
\[
\eta=\frac{\omega}{\omega_L}\frac{p_{N_0+1}^{(\infty)}-p_{N_0}^{(\infty)}}{p_{N_0+1}^{(\infty)}-p_{N_0+1}^{(0)}}.
\]
Evaluated for a cavity that starts in the thermal state $e^{-\beta H_{\rm cavity}}/Z_{\rm cavity}$ gives (remember $\omega<\omega_L$)
\[
\eta^{\omega_\beta}=\frac{\omega}{\omega_L}\frac{(1-e^{-\beta\hbar(\omega_e-\omega_g)})(1+e^{-\beta\hbar\omega})}{1-e^{-\beta\hbar \omega_L}}=\frac{\omega}{\omega_L}\left(1-\frac{e^{-\beta\hbar \omega_L}e^{\beta\hbar\omega}-e^{-\beta\hbar\omega}}{1-e^{-\beta\hbar\omega_L}}\right)\leq 1.
\]
Charging from the passive state of Eq.~\eqref{chselect} i.e. $p_{N_0}^{(0)}=p^{(\rm th)}_g,p_{N_0+1}^{(0)}=p^{(\rm th)}_e$, the efficiency is $\eta^{\sigma_{\omega_\beta}}=\omega/\omega_L\leq 1.$ Comparing both efficiencies we observe that $\eta^{\omega_\beta}>\eta^{\sigma_{\omega_\beta}}$ when $\omega<\omega_L/2$ and $\eta^{\omega_\beta}<\eta^{\sigma_{\omega_\beta}}$ when $\omega>\omega_L/2$ but they approach each other as the temperature decreases. 

Let us analyze the charging rate. The eigenvalues of the matrix in Eq.~\eqref{mcavity} are 1, (associated to the invariant state) and $\frac{1}{2}(1+\cos(2G_{N_0}\tau))$ determining the rate $-\ln \frac{1}{2}(1+\cos(2G_{N_0}\tau))$ at which the probability approaches the invariant state. Considering that $G_{N_0}=(|\Omega_L g^*|/\delta)\sqrt{N_0+1}$ and the restriction for the validity of this theory, $\delta\gg |g|\gg |\Omega_L|$, to achieve a large charging rate one has to increase $|g|$ according to the constraint. 

For realistic values of the parameters, the equilibrium state is metastable, in the sense that the closer the parameters are to the selective regime, the longer the population is in the $p_{n}^{(\infty)}$ state. We show this numerically. 

\begin{figure}[h] 
\centering
\includegraphics[width=0.4\textwidth]{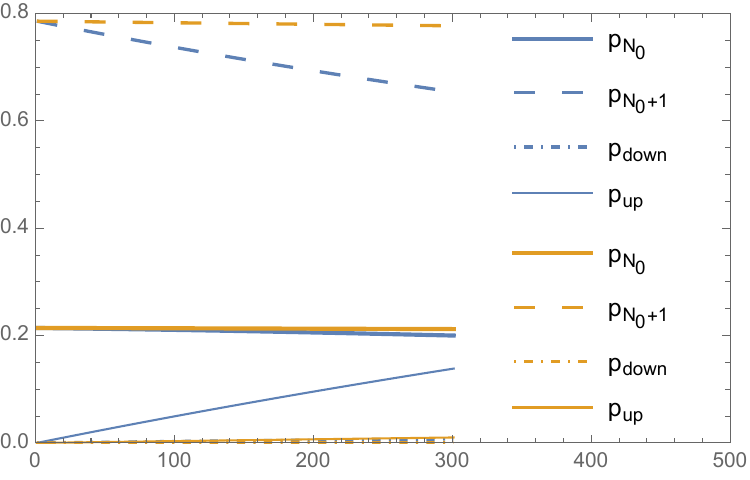}
\includegraphics[width=0.4\textwidth]{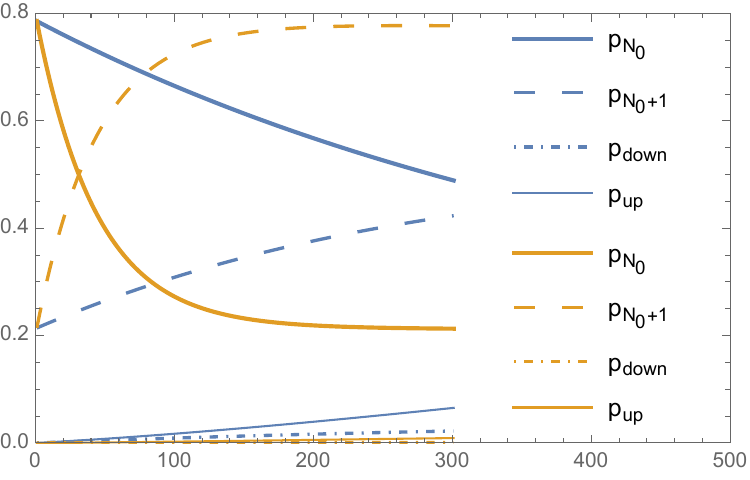}
    \caption{Fixed in all plots of the figure: $\tau=1.0, N_0=5, \Omega_L=0.1, \delta=94., \epsilon=1.3,\beta=1.$ (Left panel) initial condition: Cavity in the steady state of the selective limit $p^{(\rm th)}_e\ket{N_0}\bra{N_0}+p^{(\rm th)}_g\ket{N_0+1}\bra{N_0+1}$ in blue $g=18$ and in orange $g=58$. (Right panel) initial condition cavity in $p^{(\rm th)}_g\ket{N_0}\bra{N_0}+p^{(\rm th)}_e\ket{N_0+1}\bra{N_0+1}$. Obs: $p_{\rm up}=\sum_{n=N_0+2}^\infty p_n$ and $p_{\rm down}=\sum_{n=0}^{N_0-1} p_n$}
    \label{figura qubit}
\end{figure}

To illustrate our results, we evolve numerically the probability distribution ${\bf p}^{(i)}$ according to Eq.~\eqref{tridiag} and plot it in figure~\ref{figura qubit} as a function of $i$.  In the right panel of figure~\ref{figura qubit}, we start with a normalized probability distribution in the sector $\{N_0,N_0+1\}$ but without population inversion and observe that population inversion is achieved faster as $g$ is greater. We also observe that the population of the levels $n>N_0+1$ slowly increases. This illustrates the fact that there is no stationary distribution. But as follows from our previous discussion, the population of these levels is negligible in the high selectivity limit. To see this more clearly, we plot in the left panel of figure~\ref{figura qubit} the evolution starting with the steady state solution in the high selectivity limit 
$\rho_{\rm cavity}=(1-p^{(\rm th)}_g)\ket{N_0}\bra{N_0}+p^{(\rm th)}_g\ket{N_0+1}\bra{N_0+1}$ as initial condition. 
We see that the system stays closer to this state as $g\gg \Omega_L,$ and  $\delta\gg g, \Omega_L$.

\begin{figure}[h] 
\centering
\includegraphics[width=0.4\textwidth]{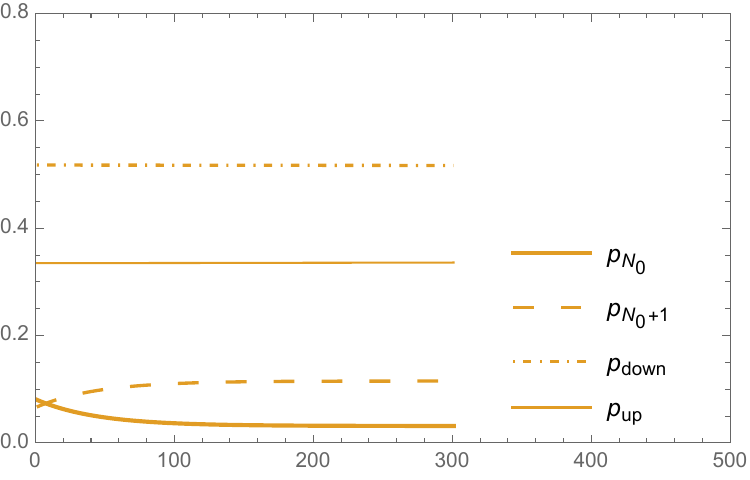}
\includegraphics[width=0.4\textwidth]{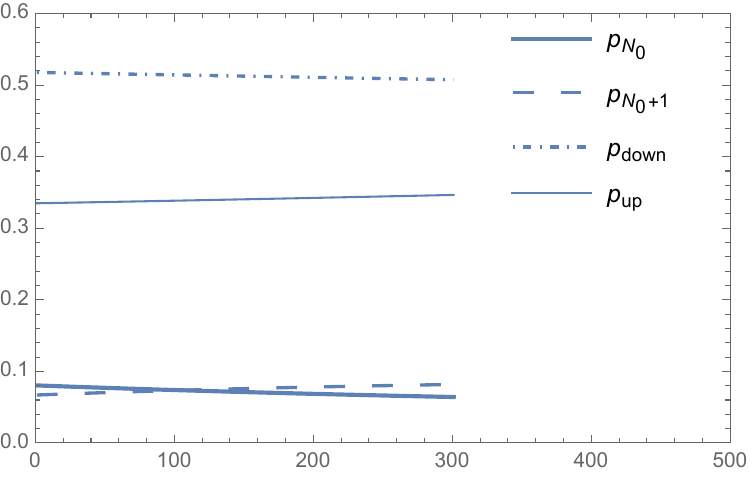}
    \caption{Fixed in all plots of the figure: $\tau=1.0, N_0=5, \Omega_L=0.1, \delta=94., \epsilon=1.3,\beta=1.$ Initial condition: Cavity in a thermal state with frequency such that $\bar{n}_{\rm photon}=5$. Blue (right panel) $g=18$ and orange $g=58$ (left panel). For the orange case ($g=58$) where the state is metastable, the ratio of $p_{N_0+1}/p_{N_0}\to e^{\beta \hbar(\omega_e-\omega_g)}$. }
    \label{figura2}
\end{figure}

In figure \ref{figura2}, we consider the evolution Eq.~\eqref{tridiag} with a thermal initial condition for the cavity. The left panel shows that the population of $N_0$ and $N_0+1$ are quickly inverted and that the other populations remain invariant (the parameters are in the high selectivity limit). The right panel departs from this limit, and we observe that the population of $N_0$ and $N_0+1$ are inverted at a slower pace, but more importantly, the population of the higher levels starts to increase.

\section{Conclusions}\label{secconclu}

We have presented two related and complementary quantum batteries that rely on the interaction between a three-level atom and a QED cavity. The charging processes are robust and almost autonomous. In the first, the battery is the atom that crosses a sequence of thermal cavities to get charged. In the second, the cavity gets charged after a stream of atoms crosses it. Although both batteries rely on similar physical setups, they present important differences. For instance, we have seen that, to increase the charging rate of the qubit battery, one needs to increase the laser intensity $\Omega_L$ to increase $A_\tau$, while to increase the charging rate of the cavity battery, one needs to increase the coupling with the cavity $g$. Another difference is that the atom always reaches the charged equilibrium state, whereas the cavity does not reach equilibrium unless the conditions for high selectivity are satisfied. In both cases, the battery is charged if one considers a sufficient number of iterations of the map.

We have characterized the charge of the batteries by its ergotropy, but the main physical mechanism is to produce a population inversion in the battery's state by interacting with thermal systems. Similar batteries have been considered in~\cite{felipe1,felipe2, marcelo2, marcelo3}, and their interest resides in the fact that the environment is not detrimental to their storing capacity~\cite{felipe3}.  

The model here described can be applied to any physical system operating in the so-called strong coupling regime, i.e., when the coupling strengths $\Omega_L,g$ are large enough when compared to decoherence rates so that the effective anti-Jaynes-Cummings doublets can be spectrally resolved. Examples of many different systems that reach this regime are summarized in~\cite{strongcouplingbook}. Just as an example, for Rydberg atoms interacting with high-Q microwave cavities, $g/2\pi\sim 50$ kHz~\cite{strongcouplingbook}. For $N\sim 10$, $\delta/2\pi\sim 1$MHz and $\Omega_L\sim g/30$, the typical time $\tau$ will be of the order of 1 ms, whereas decoherence times can be as high as 1s nowadays. Thus, a stream of thermal Rydberg atoms and a maser with a frequency close to a Rydberg atomic transition ($\omega_L\sim 50$ GHz) will charge the microwave cavity.

\section*{acknowledgments}
Z.B. and F.B. acknowledge support from Fondecyt project 1231210 and ANID – Millennium Science Initiative Program-NCN19-170. M.F.S. acknowledges CNPq Project 302872/2019-1 and FAPERJ project E-26/200.307/2023.

\end{document}